\documentclass[aps,prd,twocolumn,showpacs,nofootinbib,amsmath,amssymb,floatfix,superscriptaddress,showkeys]{revtex4}
\usepackage{graphicx}
\usepackage{dcolumn}
\usepackage{bm}
\usepackage{captcont}
\usepackage{xcolor}

\usepackage{epstopdf}
\usepackage{mathtools}
\usepackage{natbib}
\usepackage{ulem}
\usepackage[colorlinks,linkcolor=red,anchorcolor=blue,citecolor=blue]{hyperref}

\newcommand{\jcap}{J. Cosmol. Astropart. Phys.}

\newcommand{\physrep}{Phys. Rep.}
\newcommand{\mnras}{Mon. Not. R. Astron. Soc.}

\newcommand{\aap}{Astron. Astrophys}

\newcommand{\app} {Astropart. Phys.}

\begin{document}
\title{Limits on Dark Matter annihilation cross sections to gamma-ray lines with subhalo distribution in N-body simulation and Fermi LAT data}
\author{Yun-Feng Liang}
\affiliation{Key Laboratory of Dark Matter and Space Astronomy, Purple Mountain Observatory, Chinese Academy of Sciences, Nanjing 210008, China}
\affiliation{School of Astronomy and Space Science, University of Science and Technology of China, Hefei, Anhui 230026, China}
\author{Zi-Qing Xia}
\affiliation{Key Laboratory of Dark Matter and Space Astronomy, Purple Mountain Observatory, Chinese Academy of Sciences, Nanjing 210008, China}
\affiliation{School of Physics, University of Science and Technology of China, Hefei, 230026, China}
\author{Kai-Kai Duan}
\affiliation{Key Laboratory of Dark Matter and Space Astronomy, Purple Mountain Observatory, Chinese Academy of Sciences, Nanjing 210008, China}
\affiliation{University of Chinese Academy of Sciences, Beijing, 100012, China}
\author{Zhao-Qiang Shen}
\affiliation{Key Laboratory of Dark Matter and Space Astronomy, Purple Mountain Observatory, Chinese Academy of Sciences, Nanjing 210008, China}
\affiliation{University of Chinese Academy of Sciences, Beijing, 100012, China}
\author{Xiang Li}
\affiliation{Key Laboratory of Dark Matter and Space Astronomy, Purple Mountain Observatory, Chinese Academy of Sciences, Nanjing 210008, China}
\author{Yi-Zhong Fan}
\affiliation{Key Laboratory of Dark Matter and Space Astronomy, Purple Mountain Observatory, Chinese Academy of Sciences, Nanjing 210008, China}
\affiliation{School of Astronomy and Space Science, University of Science and Technology of China, Hefei, Anhui 230026, China}

\date{\today}

\begin{abstract}
In this work, we simulate a set of realizations of local volume dark matter subhalo population based on the distributions and relations derived from Via Lactea II N-body simulation.
We calculate the J-factors of these subhalos, and find that the low mass subhalos contribute a lot to the total J-factors.
Combining with 91 months of the Fermi LAT observation, we constrain on the cross section of dark matter annihilating directly to two gamma rays.
This is the first work combining numerical simulation results and Fermi LAT observations to constrain dark matter cross section to gamma-ray line with subhalo population.
Though the constraints derived from subhalo population are weaker than those from Fermi LAT observation of the Galactic center, they are supports of and complementary to these other results.
\end{abstract}

\maketitle

\section{Introduction}
According to the standard $\Lambda$CDM cosmology, non-baryonic cold dark matter (DM) is supposed to compose of approximately 27\% of the total energy density of the current Universe \cite{planck15cosmosParams}.
However, the existence of this mystery material is only supported by gravitational phenomena so far.
It is badly needed to discover non-gravitational evidence to prove its existence, to clarify its characteristics.
Among numerous theoretical particles proposed to interpret DM, weakly interacting massive particles (WIMPs) are the leading candidates, since they can naturally match current DM abundance supposing they are frozen out from an equilibrium state with high temperature and high density in the early universe. If DM constitute of WIMPs, owing to their weak interactions, they can generate the Standard Model particles, such as gamma rays, cosmic rays and neutrinos, by annihilation or decay.
Thus we can study the DM properties through the measurements of these secondary productions from DM annihilation or decay in regions of high dark matter density.
In fact great efforts have been made to search for such secondary productions in the past several decades \cite{bertone05review,fan10dmreview,charles16sensitivity,gaskins16dmreview}.

Among all kinds of possible DM annihilation signals, monochromatic gamma-ray lines, generated by DM annihilating to double photons directly, are of special importance.
Since no known astrophysical process could generate a line-like signal, it can be clearly discriminated from astrophysical backgrounds.
Thus once such a line signal is reliably discovered in the GeV-TeV band, it is a smoking-gun that there is something associated with new physics.
Due to the importance of the line emission, previously many works have tried to search for such signals \cite{pullen07EGRETline,bringmann12_130gev,weniger12_130gev,su2012line,geringer12dsphLine,tempel12_130gev,huang12_130gev,fermi13line,albert14line,fermi15line,anderson16gclsLine,liang16gclsLine,liang16dsphLine}.
Though most of these works did not find any signal and give only upper limits on the gamma-ray line fluxes, some of them showed indications of potential weak signals \cite{bringmann12_130gev,weniger12_130gev,su2012line,liang16gclsLine}.

The flux of a gamma-ray line generated by DM annihilating to double photons can be described by
\begin{equation}
\Phi(E)=\frac{1}{4\pi}\frac{\langle\sigma{v}\rangle_{\gamma\gamma}}{2m_{\rm DM}^2}2\delta(E-E_\gamma)\cdot{J_{\rm anni}},
\end{equation}
where $\langle\sigma{v}\rangle_{\gamma\gamma}$ is the annihilation cross section averaged by the DM velocity, $m_{\rm DM}$ is the rest mass of the DM particle, and $\delta(E-E_\gamma)$ is the delta function which denotes the spectrum of line signal in each annihilation.
The last term $J_{\rm anni}$ is related to the DM distribution in the space, often called J-factor, which is the integration of the square of DM density $\rho^2(r)$ along the line of sight $s$,
\begin{equation}
J_{\rm anni} = \int_\Omega\int_{s=0}^\infty{\rho^2(r(s))}{\rm d}s{\rm d}\Omega.
\label{jfactor}
\end{equation}
Hence the flux of DM annihilation signal is proportional to the square of DM density $\rho^2(r)$.  Thus the gamma-ray emissions from dark matter are preferentially detected from the regions with high DM density, such as the Galactic center, dwarf spheroidal galaxies, and galaxy clusters.
In the past decade, numerical N-body simulations have revealed that dark matter halos form hierarchically. As a result, a DM halo hosting the Milky Way would also host large numbers of dark matter subhalos (DMSH).
These DMSHs are also suitable targets to hunt for DM annihilation signals. So, some studies dedicated to search for DM signals from DMSHs have been performed \cite{fermi12dmsh,bertoni15dmsh,schoonenberg16dmsh,bertoni16j2212,hooper16dmsh,calore16dmsh,wyp16dmsh,xzq16dmsh}. Some high-latitude unidentified gamma-ray sources spectrally compatible with DM annihilating to $b\bar{b}$ have been picked up from fermi 3FGL catalog \cite{fermi15_3fgl}, and are treated as DMSH candidates \cite{bertoni15dmsh}. Two sources among them show evidences of spatial extension, thus are attractive candidates of DMSHs \cite{bertoni16j2212,wyp16dmsh,xzq16dmsh} though the possibility that these two sources actually consisted of two or more unresolved point sources can not be excluded, yet.
All these previous works focused on continuous signals of DM annihilating to, for example, $b\bar{b}$ or $\tau^+\tau^-$. In this work, we present our studies concerning on DM line signals from DMSHs. As far as we know, this is the first work combine numerical simulation results and Fermi-LAT observations to constrain DM cross section to gamma-ray line with DMSHs.

\section{DMSH distribution in the local volume of the Milky Way}
\label{sec_sim}

\begin{figure*}[!htb]
\includegraphics[width=0.45\textwidth]{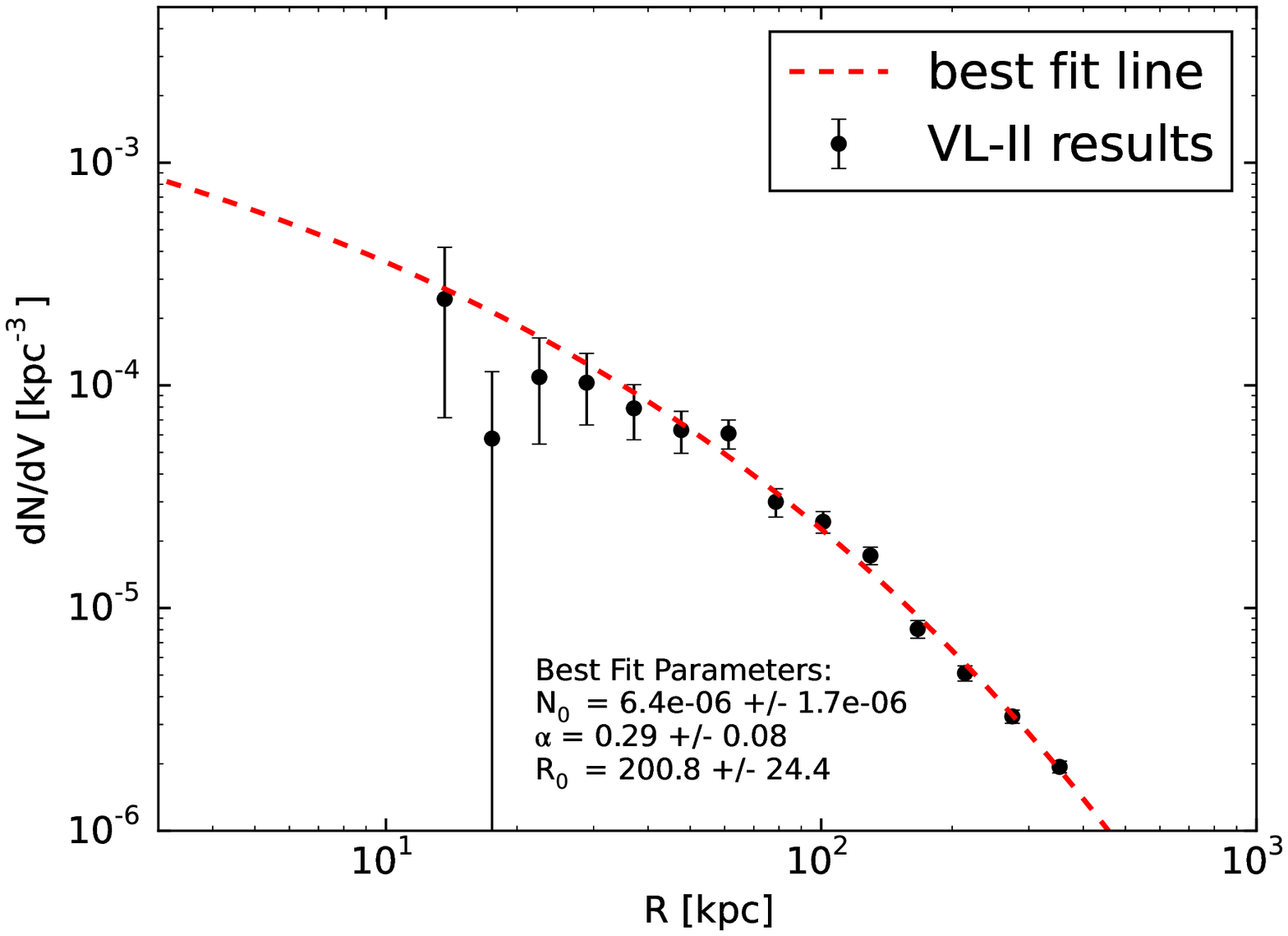}
\includegraphics[width=0.45\textwidth]{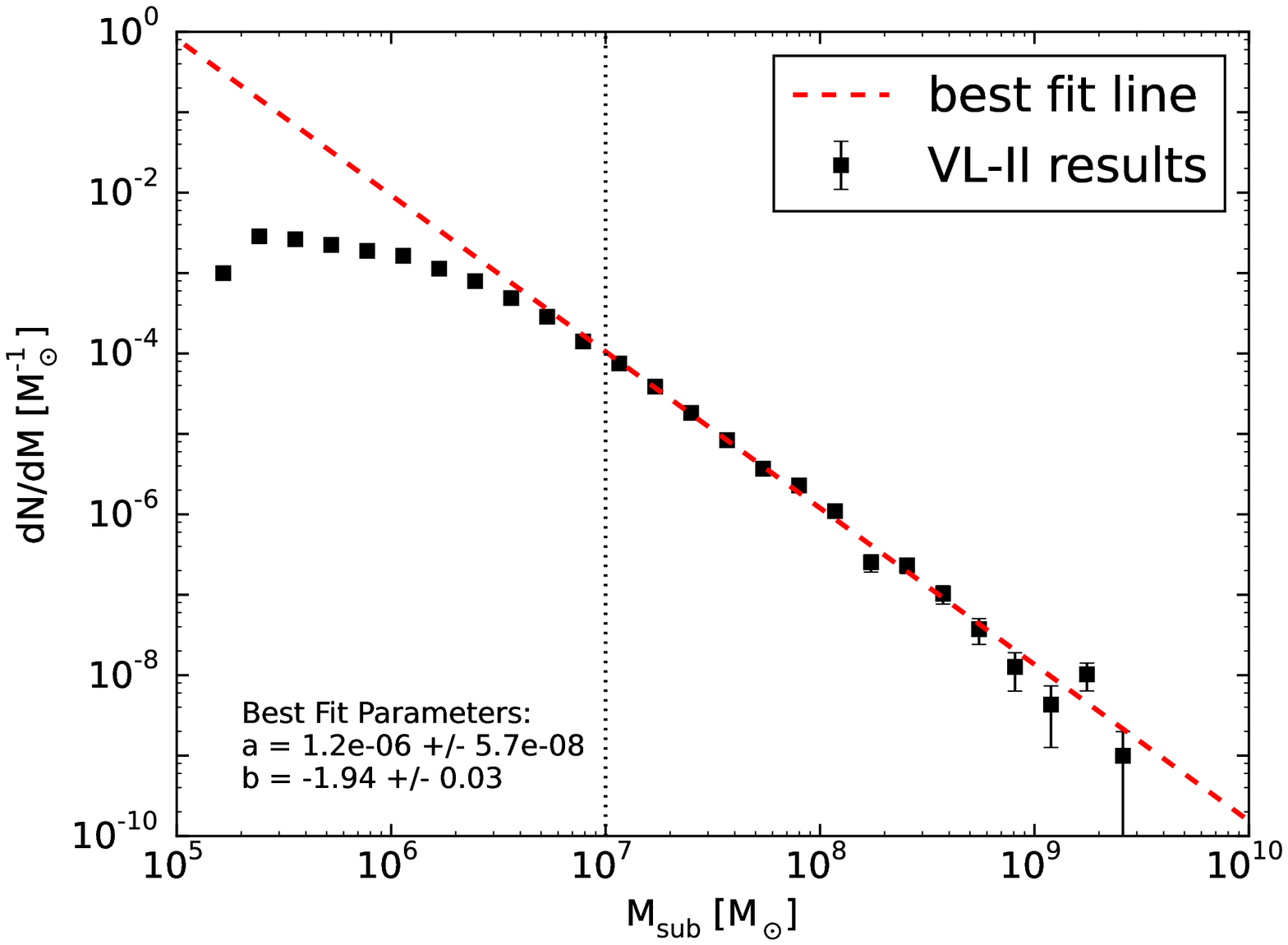}
\caption{{\it Left panel:} The radial distribution of DMSHs in VL-II simulation in the range $M_{\rm sub}=[10^7,10^{10}]~{\rm M_{\odot}}$ and $R=[3,400]$ kpc. For this distribution, we use an Einasto function to fit the data. Since the number of DMSHs below 20 kpc may have been underestimated due to resolution limitation of VL-II \cite{diemand08VL2}, the fitting is performed in a sub range of $R=[20,400]$ kpc. {\it Right panel:} The mass distribution of DMSHs in the range $R=[3,400]$ kpc. A power law function can well fit the data above $10^{10} {\rm M_\odot}$. Below this mass, VL-II data may suffer incompleteness problem. For both plots, the red dashed lines represent the best fitted lines, with its parameters listed in the plots.}
\label{fig_distri}
\end{figure*}

In the recent years, the development of the N-body simulations, such as Via Lactea II (VL-II) \cite{diemand08VL2}, Aquarius \cite{springel08Aquarius} and ELVIS \cite{garrison14ELVIS}, have significantly improved our understandings of DM distributions in the Milky Way like DM halos. All these simulations resolved numerous substructures and give very similar abundance, distribution, and other properties of DMSH population. Calore et al.\cite{calore16dmsh} also pointed out that inclusion of baryonic matters in the numerical simulations wouldn't affect the resulted gamma-ray DM annihilation signals considerably.
Therefore, for simplicity, we will only use Via Lactea II results\footnote{\url{http://www.ucolick.org/~diemand/vl/data.html}} to derive the characteristics of DMSH population.

VL-II resolved over 20,000 DMSHs which have a peak circular velocity $V_{\rm max}>4 {\rm km/s}$. In this work we only consider the subhalos located within galactocentric distance $R<400$ kpc (roughly the virial radius of the main halo). Since $R<3$ kpc is outside of our ROI (see section \ref{sec_fermi}), we also don't consider DMSHs in this region.
Our goal in this work is to estimate the J-factor of DMSH population in the local volume of the Milky Way, and then use it to constrain on DM cross section. To estimate the J-factor, we first derive analytical DMSH spatial and mass distributions by analyzing the results of VL-II. And then using these analytical descriptions, we simulate 500 Monte Carlo realizations of DMSH populations in the local volume of the Milky Way. For each subhalo in the mock data, we sample its spatial position, mass and internal DM distribution to calculate its J-factor. At last, the total J-factor of a DMSH population is determined by summing over all individual DMSH. We don't use the VL-II subhalos directly for the following reasons. VL-II resolved subhalos down to $10^5M_\odot$, however, at masses below $10^7M_\odot$ it will suffer a completeness problem due to the finite resolution of the simulation. Moreover, theoretical arguments predict Milky Way subhalos can still exist down to a minimal mass of $10^{-6}M_{\odot}$ \cite{fermi12dmsh,hofmann01minDMSHmass,loeb05minDMSHmass}. In the inner tens of kiloparsecs, VL-II results also suffer the same incompleteness issue due to simulation resolution. Thus, to consider these lower mass and inner region populations an extrapolation is required. When extracting the analytical position and mass distribution of DMSH, we restrict our analysis in the range $M_{\rm sub}\in[10^{7},10^{10}]~{\rm M_\odot}$ and $R\in[3,400]$ kpc.

\begin{figure*}[!htb]
\includegraphics[width=0.45\textwidth]{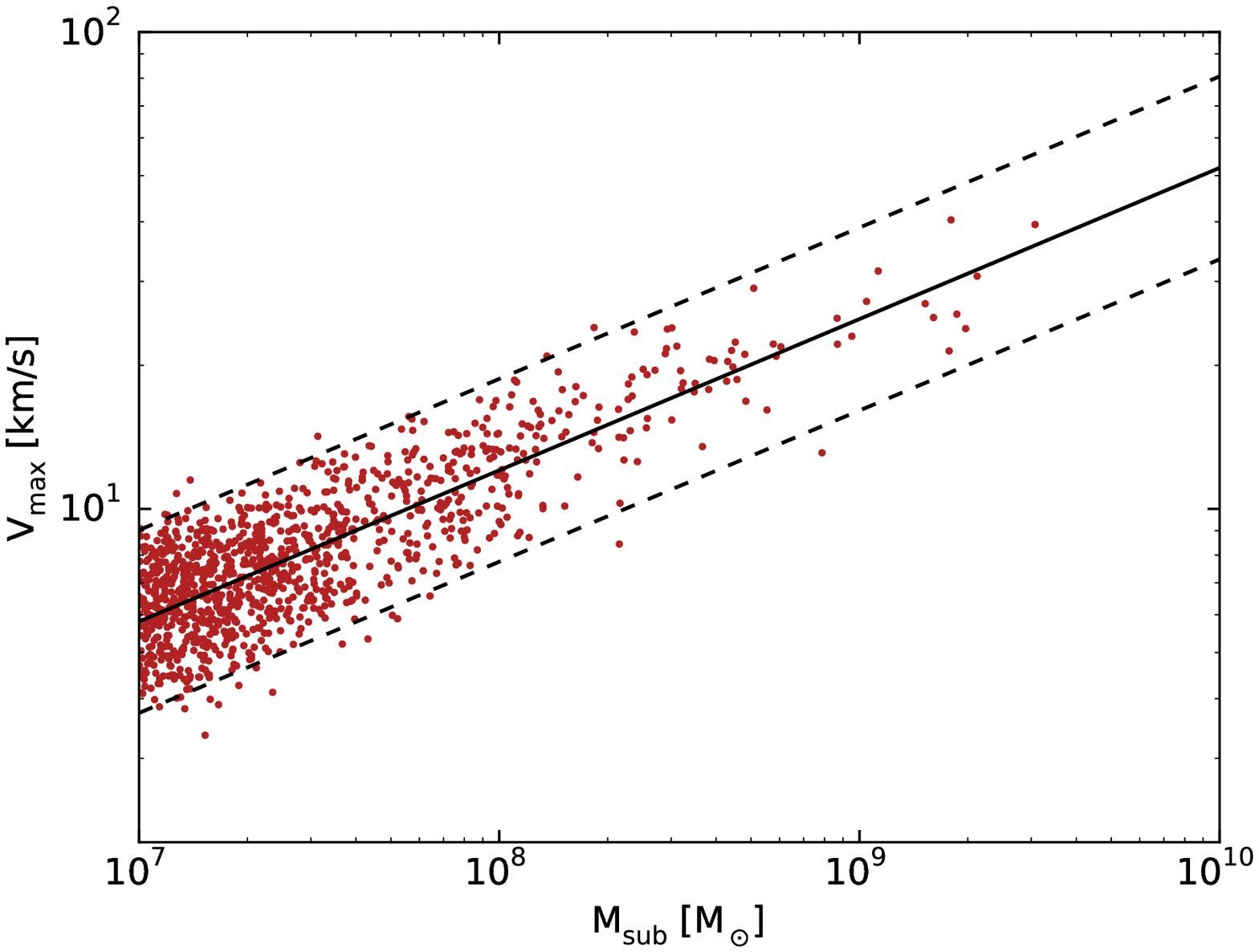}
\includegraphics[width=0.45\textwidth]{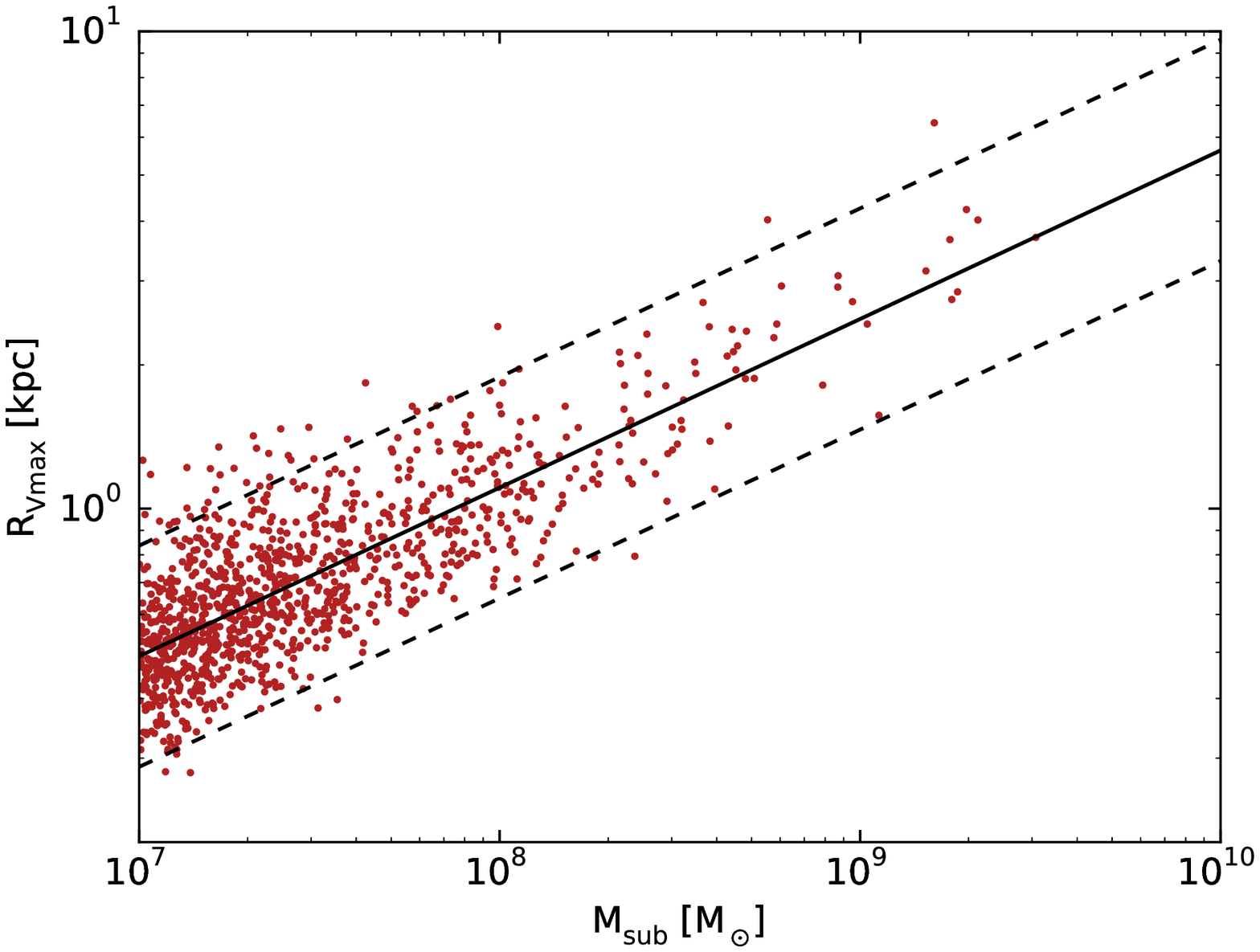}
\caption{Correlation between the maximum circular velocity of subhalos $V_{\rm max}$ (the radius at which $V_{\rm max}$ is reached, $R_{V_{\rm max}}$) and the subhalo mass $M_{\rm sub}$. The dashed lines represent the $2\sigma$ scatter of these correlations.}
\label{fig_relation}
\end{figure*}

The left panel of Fig.\ref{fig_distri} shows the radial distribution of VL-II subhalos in the ranges of $M_{\rm sub}=[10^7,10^{10}]~{\rm M_{\odot}}$ and $R=[3,400]$ kpc.
We adopt an Einasto function \cite{springel08Aquarius}
\begin{equation}
\frac{dN}{dV}(R)=N_0{\rm exp}\{-{(\frac{2}{\alpha})[(\frac{R}{R_0})^\alpha-1]}\}
\end{equation}
to fit the radial distribution data in Fig.\ref{fig_distri}. Considering that the true subhalo abundance inside 20 kpc may be underestimated due to resolution limitations \cite{diemand08VL2}, the fit is performed in a sub range of $R>20$ kpc.
The dashed red is the best fit line adopting a maximum likelihood fit, with the optimal parameters presented in the plot.

For the mass distribution, a power law function
\begin{equation}
\frac{dN}{dM}(M)=a{\cdot}(\frac{M}{10^8{\rm M_\odot}})^b
\end{equation}
can fit the VL-II data well for masses above $10^7 M_\odot$. Below this mass, the VL-II data departing from this power-law may be due to the finite resolution of the simulation. The maximum likelihood fit gives $b\sim-1.9$, well consistent with previous works \cite{fermi12dmsh,hooper16dmsh,calore16dmsh}.

We assume that the DM spatial distribution in the individual subhalo follows an NFW density profile \cite{navarro97nfw}
\begin{equation}
\rho(r)=\frac{\rho_s}{(r/r_s)(1+r/r_s)^2}.
\end{equation}
As the result of tidal stripping, the outer part of a subhalo would be removed. Considering this effect, Hooper et al.\cite{hooper16dmsh} suggests an exponential cutoff power law profile accommodates simulation results better. However, we ignore this effect in our work. The resulted bias on the predicted annihilation flux would be less than 10\% \cite{fermi12dmsh}.
To absolutely determine an NFW profile for each subhalo, the value of scale radius $r_{\rm s}$ and scale density $\rho_{\rm s}$ are needed.  These two quantities can be derived from the maximum circular velocity of subhalos, $V_{\rm max}$, and the radius at which this velocity is reached, $R_{V_{\rm max}}$, according to following expressions \cite{fermi12dmsh,kuhlen08}
\begin{equation}
\label{eq_rs1}
r_s=\frac{R_{V_{\rm max}}}{2.163}
\end{equation}
\begin{equation}
\label{eq_rhos1}
\rho_s=\frac{4.625}{4\pi{G}}\left(\frac{V_{\rm max}}{r_s}\right)^2
\end{equation}
While for $V_{\rm max}$ and $R_{V_{\rm max}}$, previous studies have shown that they are correlated with the subhalo mass. Thus for each subhalo in the MC simulation, its $V_{\rm max}$ and $R_{V_{\rm max}}$ are sampled base on these correlations which can be extracted by fitting to the VL-II data. We perform a fit to the correlation between $R_{V_{\rm max}}$ data and subhalo mass with a power law function
\begin{equation}
R_{V_{\rm max}}=R_0\cdot(\frac{M}{10^7{\rm M_\odot}})^{\delta}
\end{equation}
and derive the best-fit parameters to be $R_0=0.49$ km, $\delta=0.35$, and a log-normal scatter of $\sigma=0.116$.
For $V_{\rm max}$, there is
\begin{equation}
V_{\rm max}=V_0\cdot(\frac{M}{10^7{\rm M_\odot}})^{\beta}
\end{equation}
with $V_0=5.8$ km/s, $\delta=0.32$, and a log-normal scatter of $\sigma=0.096$. Our results are presented in Fig.\ref{fig_relation}. Similar results have already been obtained in \cite{fermi12dmsh}.

{The above approach assumes that the internal DM distribution of the lower mass subhalos ($M<10^7 M_\odot$) can be derived by the power law extrapolation of those larger ones, however some studies show that the extrapolation to low mass range is not a simple power law \cite{sc14cm,moline17bf}. In view of such a fact we also adopt another way to determine the internal DM distribution of each subhalo. The subhalo's concentration parameter $c={R_{\rm virial}}/{r_s}$ is correlated its mass $M$, as found in the N-body simulations covering more than 20 orders of magnitude in DM halo mass. With this concentration-mass (c-M) relation, one can obtain}
\begin{equation}
\label{eq_rs2}
r_s = (\frac{3M}{4{\pi}c^3{\cdot}{200}\rho_c})^{1/3}
\end{equation}
\begin{equation}
\label{eq_rhos2}
\rho_s = \frac{M}{4{\pi}r_s^3[{\rm ln}(1+c)-{c}/(1+c)]}
\end{equation}
{where $\rho_c$ is critical density of the universe, and for $c=c(M)$ we adopt the relation in \cite{sc14cm}. The scatters in c-M relation are not considered in our simulation. Once the mass of a subhalo is given, its DM density profile is also absolutely determined. In the later analysis, we will also present results based on this prescription as a comparison.}

With these distributions and relations in hand, we generate 500 Monte Carlo realizations of subhalo population.
For each realization, a set of subhalos together with their masses, positions in the Milky Way, $R_{V_{\rm max}}$, $V_{\rm max}$, $r_s$, $\rho_s$ are randomly sampled according to above distributions and relations. For the subhalos outside the ranges of $M_{\rm sub}=[10^7,10^{10}]~{\rm M_{\odot}}$ and $r=[3,400]$ kpc, we assume that they also follow the same distributions and relations thus we use an extrapolation to sample their characteristics. The total number of subhalos in each realization would be
\begin{equation}
N_{\rm tot}=\frac{N(M_{\rm min})}{N(10^7)}\int_3^{400}\frac{dN}{dV}dr
\end{equation}
\begin{equation}
N(m){\equiv}N(m<M_{\rm sub}<10^{10},20<r<400)=\int_m^{1e10}\frac{dN}{dM}dM
\end{equation}
where $M_{\rm min}$ is the minimal mass of the subhalo. We will adopt some different $M_{\rm min}$ in our later studies to account for the uncertainty of this value.

\section{J-factor of subhalo population}
\label{sec_jf}
Once the position and density profile of each subhalo is determined, we calculate its J-factor using Eq.\ref{jfactor}. For subhalos at large distances ($D>2r_{\rm tidal}$), we use the expression
\begin{equation}
J=\frac{1}{D^2}\int_0^{r_{\rm tidal}}4\pi{r^2}\rho^2(r){\rm d}r
\end{equation}
to approximate the J-factor calculation, where $D$ is the distance of the subhalo and we choose to integrate out to the tidal radius $r_{\rm tidal}$ of the subhalos. The total J-factors of SH populations are derived by summing all subhalos together. In the table \ref{tb1}, we summarize the predicted numbers and J-factors for subhalo populations in different mass range.
{The {\it J-factor1} is calculated by using $R_{V_{\rm max}}$ and $V_{\rm max}$ to derive subhalos' $r_s$ and $\rho_s$, while {\it J-factor2} utilizes the c-M relation. Generally, the later would result in smaller J-factor of subhalo population (see \cite{moline17bf} for similar conclusion). The uncertainties on J-factors correspond to 68\% variation in 500 MC simulations.}
Note that for subhalos in the mass range of $M_{\rm sub}=[10^0,10^{1}]~{\rm M_{\odot}}$ and $M_{\rm sub}=[10^1,10^{3}]~{\rm M_{\odot}}$, we only sample $10^7$ subhalos rather than all predicted number to reduce the expense of calculation, so the scatters should have been amplified.
From table \ref{tb1},
we find that there exist large numbers of low mass subhalos. Though the J-factor of each of them is small, in total they contribute a lot to the summed J-factor, even greater than those high mass populations.

\begin{figure}[!htb]
\includegraphics[width=0.45\textwidth]{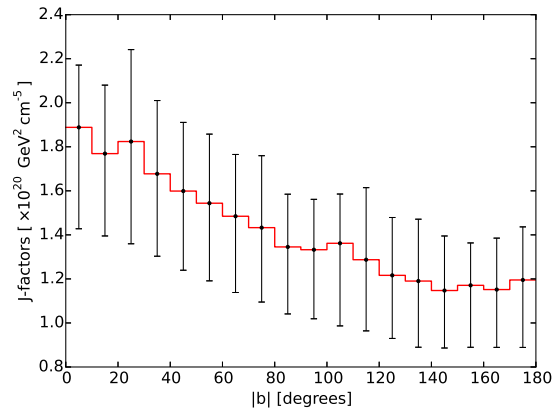}
\caption{J-factors of subhalo populations located in different longitude ranges. The error bar indicates $1 \sigma$ uncertainty in 500 MC simulations.}
\label{fig_jfs_lon}
\end{figure}

We also compare the J-factors of subhalos locate in different longitudes, which is shown in figure \ref{fig_jfs_lon}. The error bar indicates $1 \sigma$ uncertainty in 500 MC simulations. We find that the J-factor is only slightly dependent on the longitude. Thus we don't apply any longitude cut when defining our ROI in section \ref{sec_fermi}.

\begin{table}[!htb]
\caption{The total numbers and J-factors of subhalo populations in different mass range. }
\begin{ruledtabular}
\begin{tabular}{cccc}
  Mass Range & Subhalo Number  &  J-factor1\footnote{J-factors calculated by using Eq.\ref{eq_rs1} and \ref{eq_rhos1} to derive subhalos' $r_s$ and $\rho_s$.} &  J-factor2\footnote{J-factors calculated by using Eq.\ref{eq_rs2} and \ref{eq_rhos2} to derive subhalos' $r_s$ and $\rho_s$.}\\
 $[M_\odot]$ &     &\multicolumn{2}{c}{[$\times10^{20}~$GeV$^2$cm$^{-5}$]} \\
 \hline
 $[10^0$,$10^1]$    &  $4.02\times10^9$ & $3.7_{-0.2}^{+0.4}$  & $1.0_{-0.1}^{+0.1}$  \\
 $[10^1$,$10^3]$    &  $5.09\times10^8$ & $6.6_{-0.6}^{+1.0}$  & $1.8_{-0.1}^{+0.2}$  \\
 $[10^3$,$10^5]$    &  $6.56\times10^6$ & $5.8_{-0.4}^{+0.7}$  & $1.4_{-0.1}^{+0.1}$  \\
 $[10^5$,$10^7]$    &  $8.47\times10^4$ & $5.1_{-0.4}^{+0.7}$  & $1.1_{-0.1}^{+0.1}$  \\
 $[10^7$,$10^{10}]$ &  $1.11\times10^3$ & $5.5_{-1.2}^{+2.9}$  & $1.1_{-0.2}^{+0.6}$  \\
 Total              &  $4.54\times10^9$ & $26.7_{-2.0}^{+3.7}$ & $6.4_{-0.4}^{+0.8}$ \\
\end{tabular}
\end{ruledtabular}
\label{tb1}
\end{table}


\section{Analyzing the Fermi LAT data}
\label{sec_fermi}
The Fermi-LAT \cite{atwood09lat} is currently one of the most prominent gamma-ray instruments on-orbit, which is sensitive in the energy range $\sim$30~MeV to {\textgreater}500~GeV and most appropriate for indirect searches of WIMPs.
In this work, we will use the publicly released Fermi-LAT Pass 8 data (P8R2 Version 6) to perform our analysis. One of the improvements of the Pass 8 data is that they can be subdivided into quartiles according to events' energy/direction reconstruction qualities, allowing to improve the energy/direction resolution by using the high quality data only \cite{fermi13pass8econf}. In total, 91 months' data (from 2008-10-27 to 2016-06-08, i.e. MET 246823875 - MET 487121910) are used. We take into accont the energy range between 1 and 500 GeV.  The Fermi collaboration recommended  zenith-angle cut $\theta < 90^\circ$ and quality-filter cuts (DATA\_QUAL{\textgreater}0 \&\& LAT\_CONFIG==1)  are applied  in the {\it gtselect} and {\it gtmktime} steps. Since we are mainly interested in the high latitude regions where the residual cosmic-ray background is an important contamination, we make use of the ULTRACLEAN data. Considering the energy resolution of EDISP0 data is much worse than that of the rest, and also for consistent with our previous works \cite{liang16gclsLine,liang16dsphLine}, we will only take into account EDISP1+EDISP2+EDISP3 data in the analysis to achieve a better energy resolution without considerably losing the statistics.
The data selection and the exposure calculation procedure are performed utilizing the latest v10r0p5 version of Fermi science tools.

We define our region of interest (ROI) as all the sky except the region of galactic latitude $|b| < 20^\circ$. Masking the low latitude region is to avoid the strong gamma-ray background from the Galactic plane.
In the LAT third source catalog (3FGL) \cite{fermi15_3fgl}, there are totally 238 sources identified as coming from pulsars, blazars, supernova remnants and other astrophysical sources. These sources are reliably excluded as DMSHs and most of them emit very strong gamma-rays. Thus we remove photons that are within $1^\circ$ radius of each of these sources (for LMC and SMC, we mask $3^\circ$ regions considering their extension). In fact whether masking the identified sources or not don't affect the final results significantly since they only contribute $\sim$5\% of the total emission for the energy range 1-500 GeV. DMSHs within these masked regions have also been ignored when calculating J-factors in section \ref{sec_jf}.

To give the 95\% confidence level (CL) upper limits of a line signal, we perform unbinned maximum-likelihood fittings to the observed count spectra of the ROI. We fit the spectra in the energy domain only, ignoring the events' spatial information, since there is no knowledge of the exact direction of the subhalos. Adopting an unbinned analysis is for not losing sensitivity due to energy binning.
The unbinned likelihood function is expressed as\cite{fermi13line}
\begin{equation}
\ln\mathcal{L}(\lambda)=\sum_{\rm i=1}^N \ln{C}(E_{\rm i})- {\int}C(E;\lambda)dE,
\label{eq_like}
\end{equation}
where $N$ is the number of total observed photons, $E_{\rm i}$ is energy of each photon, and $C(E;\lambda)$ is the expected count spectra with its model parameters $\lambda$.

For a specific DM mass, the line signal due to DM annihilation covers very narrow energy range (which is still the case when the instrumental energy dispersion function has been taken into account). For example, at 40~GeV the 68\% containment of a line signal is $\sim$5\%.  Thus we can fit the spectra in a very narrow energy window. In our work, we choose the window size to be $(E_{\gamma}-0.5E_{\gamma},~E_{\gamma}+0.5E_{\gamma})$, where $E_{\gamma}$ is the energy of putative line signal (namely the mass of DM for the case of DM annihilating to two gamma-rays).
In such a small energy window, the background spectrum could approximate to a power law. Thus the $C(E)$ in Eq.(\ref{eq_like}) can be expressed as
\begin{equation}
C(E;N_{\rm b},\Gamma)=N_{\rm b}{\cdot}E^{-\Gamma}\bar{\epsilon}(E)+N_{\rm s}{\cdot}D_{\rm eff}(E;E_{\rm line}),
\end{equation}
where $N_b$ ($N_s$) is the normalization factor of background (line signal) component, $\bar{\epsilon}(E)$ is the energy-dependent average exposure in the ROI\footnote{It is derived by averaging all pixels in the HEALPIX format exposure file within the ROI.}.
The $D_{\rm eff}$ is effective energy dispersion of the data in the ROI, which can be evaluated by following expression in the case of our analysis,
\begin{equation}
D_{\rm eff}(E;E_{\rm line}) = \frac{\sum_j\int{A(E,\theta,s_j)D(E;E_{\rm line},\theta,s_j)}{\rm d}\Omega}{\sum_j\int{A(E,\theta,s_j)}{\rm d}\Omega}
\end{equation}
where $A(E,\theta,s)$, $D(E;E',\theta,s)$ are the Fermi LAT effective area and the energy dispersion function\footnote{\url{http://fermi.gsfc.nasa.gov/ssc/data/analysis/documentation/Cicerone/Cicerone_LAT_IRFs/}}, both of which are functions of the incline angle with respect to the boresight $\theta$ and event-type parameter $s$.
By maximizing the likelihood value in Eq.(\ref{eq_like}), the model parameters best describing the observed data are obtained.
At a given line energy, the 95\% confidence level upper limit of the flux of the line signal is derived by finding the value of $N_{\rm s}$ at which the log-likelihood $\ln\mathcal{L}$ is smaller by 1.35 compared to the maximum one.
For more details of the method of gamma-ray line signal searching, we refer the readers to \cite{weniger12_130gev,fermi13line,liang16gclsLine}.

The line energy $E_\gamma$ is fixed in the fitting procedure. We will calculate the upper limits for a series of $E_\gamma$ from 5~GeV to 300~GeV with increment in steps of 0.5 ${\sigma}_E(E_{\gamma})$, where ${\sigma}_E(E_{\gamma})$ is the energy resolution ($68{\%}$ containment) of the LAT at $E_{\gamma}$.

\section{Constraint on the cross section of DM annihilating to gamma-ray lines}

\begin{figure}[!b]
\includegraphics[width=0.95\columnwidth]{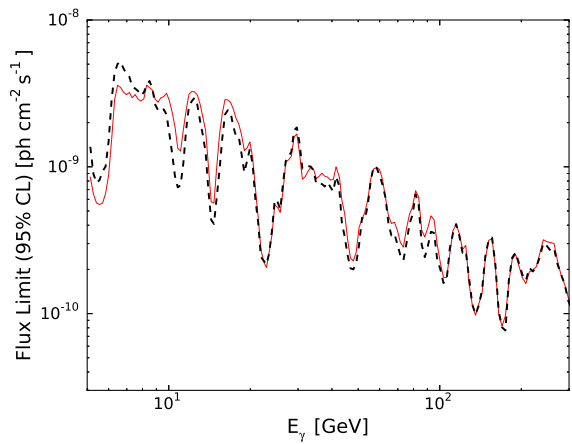}
\caption{Flux limits of line-like signals  at 95\% confidence level at energies between 5 GeV and 300 GeV for Fermi LAT 8 years' data within the ROI defined in the text. The red solid (black dashed) line is for using a power law (log parabola) function to approximate the background spectrum in each small energy window.}
\label{fig_fluxul}
\end{figure}

Fig.\ref{fig_fluxul} shows the 95\% confidence level upper limits on the gamma-ray flux of potential line signals within our ROI obtained using the analyzing method described above. Our results may be biased if the real background spectra is intrinsic curved in the adopted narrow energy windows. Thus we have also tested using a log parabola function instead of a power law one to approximate the background spectrum in each energy window. In Fig.\ref{fig_fluxul}, the red solid line is for the power law assumption and the black dashed line is for log parabola. As we can see, the log parabola assumption gives rather similar results, suggesting that in such small energy windows a power law spectrum is a good approximation to the background. Thus our subsequent results will be based on the power law assumption as that in our previous works and other line signals searching works.

\begin{figure}[!tb]
\includegraphics[width=1.0\columnwidth]{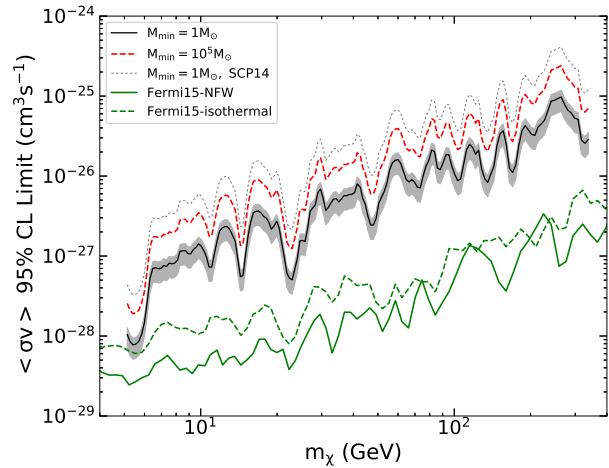}
\caption{The 95\% confidence level upper limits on the cross sections of DM annihilating into double $\gamma$-rays obtained in our analysis. The black solid line and the red dashed line are for different minimal subhalo mass assumptions. {The gray dotted line represents the upper limits derived using {\it J-factor2} and assuming a minimal subhalo mass of $1M_\odot$.} As a comparison, we also plot the constraints obtained according to the Fermi LAT observation of the inner Galactic region as green lines \cite{fermi15line}.}
\label{fig_dmul}
\end{figure}

Combining the line signal flux upper limits with the J-factors derived in section \ref{sec_jf}, we can obtain the constraints on the cross section for DM directly annihilating to two gamma rays, as shown in Fig.\ref{fig_dmul}.
The black solid line and the red dashed line represent the limits obtained assuming a minimal subhalo mass of $1 M_\odot$ and $10^5 M_\odot$, respectively.
{The gray dotted line is obtained using {\it J-factor2} (see section \ref{sec_jf} for details) and a minimal subhalo mass of $1M_\odot$ is assumed.}
All these results are derived using the mean J-factors averaged over 500 MC realizations. In the case of minimal mass of $1 M_\odot$, we also show the 95\% coverage considering the J-factors' scatter in the MC simulation (shaded band).
Contrary to searches for continuous signals from DM annihilating to quarks and leptons, for which only nearby massive subhalos can produce sufficient emission and be distinguished from the gamma-ray background, for DM line signals low mass subhalos would also contribute a lot to the total predicted signals. It is suggested in some literature that the minimal mass of DMSHs can be as low as $10^{-6}~M_\odot$, the results would be further improved if we extrapolate to such a minimal mass.

Currently, the most stringent constraints on the cross section of DM annihilating to line signals in this energy range are derived from Fermi LAT observations of the inner Galactic region.
Here we also compare our results with these constraints by \cite{fermi15line}. The green dashed and solid lines are constraints in \cite{fermi15line} for isothermal and NFW DM density profiles, respectively.
For both density profiles, the limits by DMSH populations are weaker than those derived from the Galactic data. However, our results are supports of and complementary to the these previous works.

\section{Summary}
N-body simulations reveal that there are large numbers of DMSHs in a Milky Way like DM halo. These DMSHs may generate considerable DM annihilation signals since they are regions with higher DM density compared to the smoothly distributed dark matter halo. Previous studies have examined the unidentified point sources in the Fermi-LAT catalogs, by comparing the spectra and spatial extensions of these unidentified sources with those predicted from DM annihilation, and some DMSH candidates have been suggested. Though none of these candidates can be reliably confirmed as a DMSH, stringent constraints on the DM annihilation cross section can be derived based on the number of DMSH candidates. However, all these works were concerning on continuous DM signals of DM annihilating to $b\bar{b}$ or $\tau^+\tau^-$.

In this work, we focus on using the N-body simulation results (VL-II results in our work) combining with Fermi-LAT data to constrain DM annihilation cross section to gamma-ray line signals. To do this, we first derive the characteristics, including radial distribution, mass distribution and density profile, of subhalo populations. With these derived characteristics, we generate 500 MC realization of subhalo populations in the Milky Way like DM halo and estimate the prospective J-factors of DM annihilation in subhalos.  Then we analyze Fermi-LAT Pass 8 data in the energy range of 1-500 GeV within a region of interest of $|b|>20^\circ$ to derive the 95\% CL gamma-ray flux upperlimits of potential line signals.  Utilizing the derived J-factors and gamma-ray flux upper limits, we set constraints on the cross sections of DM annihilating to line signals. Our constraints are somewhat weaker than those derived from Fermi observation of the Galactic center, however they are supports of and complementary to other constraints set by the Galactic center, dwarf galaxies and galaxy clusters. We also point out that for the line signal search, low mass subhalos would contribute a lot to the total expected annihilation flux due to their large amount. In this work, we assume the minimal subhalo masses of $1 M_\odot$ and $10^5 M_\odot$, if the subhalos can be survived down to a mass of $10^{-5} M_\odot$, as suggested in some literature, the current constraints would be further improved.

Finally, we would like to point out that an operating space mission, the Dark Matter Particle Explorer (DAMPE) \cite{chang2014dampe,chang17dampe} with an energy resolutions better than 1.5\% at energies $>$50GeV \cite{Zhang16dampeBGO}, and some proposing future missions including for instance the High Energy cosmic-Radiation Detection Facality \cite{zhang14herd} and the GAMMA-400 gamma-ray telescope \cite{galper14gamma400}, are expected to contribute significantly to the gamma-ray line search in the next decade.

\begin{acknowledgments}
We thank the referee for helpful comments. We also thank Qiang Yuan for helpful discussions. This work was supported in part by the National Basic Research Program of China (No. 2013CB837000) and  the National Key Program
for Research and Development (No. 2016YFA0400200), and National Natural Science Foundation of China under grants No. 11525313 (i.e., the Funds for Distinguished Young Scholars).
\end{acknowledgments}

\bibliographystyle{apsrev4-1-lyf}
\bibliography{subhalo_line}

\end{document}